\documentclass[conference]{IEEEtran}
\IEEEoverridecommandlockouts

\def\BibTeX{{\rm B\kern-.05em{\sc i\kern-.025em b}\kern-.08em
    T\kern-.1667em\lower.7ex\hbox{E}\kern-.125emX}}

\usepackage{tikz}
\usepackage{amsmath}

\usepackage{pifont}
\newcommand{\cmark}{\ding{51}}%
\newcommand{\xmark}{\ding{55}}%

\usepackage{textcomp}
\usepackage{xcolor}
\usepackage{hyperref}
\usepackage{graphicx}
\usepackage{booktabs} 
\usepackage{algorithmic}
\usepackage{algorithm}
\usepackage{longtable}
\usepackage{subcaption}
\usepackage{wrapfig}
\usepackage{multirow}
\usepackage{fancyvrb}
\usepackage{fancyhdr}
\usepackage{xspace,soul}
\usepackage{tcolorbox}
\usepackage{caption}
\usepackage{xcolor}
\def\mystrut{\rule{0pt}{1\normalbaselineskip}} 
\usepackage{colortbl}
\usepackage{comment}

\newcommand{\eat}[1]{}

\newcommand{\eg}{{\it e.g.,}\xspace}

\newcommand{\ie}{{\it i.e.,}\xspace}\soulregister{\ie}{7}

\newcommand{\ci}{{\it (i) }}
\newcommand{\cii}{{\it (ii) }}
\newcommand{\ciii}{{\it (iii) }}

\newcommand{\Note}{{\bf Note: }}
\newcommand{\wrt}{{\it w.r.t.}\xspace}

\newcommand{\topr}{TOPr\xspace}
\newcommand{\aflgo}{AFLGo\xspace}
\newcommand{\sfuzz}{SieveFuzz\xspace}

\newcommand{\NumBugs}{24\xspace}
\newcommand{\NumBuggyProjects}{5\xspace}
\newcommand{\hdfBugs}{13\xspace}
\newcommand{\hdfBugsHigh}{12\xspace}
\newcommand{\hdfBugsConfirmed}{12\xspace}
\newcommand{\hdfBugsFixed}{12\xspace}
\newcommand{\netcdfBugs}{5\xspace}
\newcommand{\lrzipBugs}{1\xspace}
\newcommand{\giflibBugs}{4\xspace}
\newcommand{\mjsBugs}{1\xspace}
\newcommand{\NumConfirmedBugs}{18\xspace}
\newcommand{\NumFixedBugs}{\hdfBugsFixed}

\newcommand{\mleak}{Memory Leaks}
\newcommand{\segv}{Segmentation Fault}
\newcommand{\mallocerr}{Memory Allocation Error}
\newcommand{\hbof}{Heap Buffer Overflow}

\newcommand{\mcpo}{Memcpy Parameter Overlap}

\newcommand{\tctToprAflgo}{73\%\xspace}
\newcommand{\tctToprSF}{222\%\xspace}
\newcommand{\trcToprAflgo}{9\%\xspace}
\newcommand{\trcToprSF}{149\%\xspace}
\newcommand{\ttbToprAflgo}{8\%\xspace}
\newcommand{\ttbToprSF}{85\%\xspace}

\newcommand{\Comment}[1]{}
\newcommand{\Space}[1]{}

\begin{document}


\title{TOPr: Enhanced Static Code Pruning for Fast and Precise Directed Fuzzing}


\author{\IEEEauthorblockN{Chaitra Niddodi}
\IEEEauthorblockA{\textit{Department of Computer Science} \\
\textit{University of Illinois at Urbana-Champaign}\\
USA \\
chaitra@illinois.edu}
\and
\IEEEauthorblockN{Stefan Nagy}
\IEEEauthorblockA{\textit{Kahlert School of Computing} \\
\textit{University of Utah}\\
USA \\
snagy@cs.utah.edu}
\and
\IEEEauthorblockN{Darko Marinov}
\IEEEauthorblockA{\textit{Department of Computer Science} \\
\textit{University of Illinois at Urbana-Champaign}\\
USA \\
marinov@illinois.edu}
\and
\IEEEauthorblockN{Sibin Mohan}
\IEEEauthorblockA{\textit{Department of Computer Science} \\
\textit{The George Washington University}\\
USA \\
sibin.mohan@gwu.edu}
}






\maketitle
\pagestyle{plain}

\begin{abstract}
Directed fuzzing is a dynamic testing technique that focuses exploration on specific, pre-targeted program~locations.
Like other types of fuzzers, directed fuzzers are most effective when maximizing testing speed and precision.
To this end, recent directed fuzzers have begun leveraging \emph{path~pruning}: preventing the wasteful testing of program paths deemed {irrelevant} to reaching a desired target location.
Yet, despite code pruning's substantial speedup, current approaches are {imprecise}---failing to capture indirect control flow---requiring additional dynamic analyses that diminish directed fuzzers' speeds.
Thus, without code pruning that is both fast \emph{and} precise, directed fuzzers' effectiveness will continue to remain limited.
    
This paper aims to tackle the challenge of upholding both speed and precision in pruning-based directed fuzzing.
We show that existing pruning approaches fail to recover common-case indirect control flow; and identify opportunities to enhance them with lightweight heuristics---namely, function signature matching---enabling them to maximize precision \emph{without} the burden of dynamic analysis.
We implement our enhanced pruning as a prototype, \topr (\textbf{T}arget \textbf{O}riented \textbf{Pr}uning), and evaluate it against the leading pruning-based and pruning-agnostic directed fuzzers SieveFuzz and \aflgo. 
We show that \topr's enhanced pruning outperforms these fuzzers in (1) \emph{speed} (achieving \tctToprSF and \tctToprAflgo higher test case throughput, respectively); (2) \emph{reachability} (achieving \trcToprSF and \trcToprAflgo more target-relevant coverage, respectively); and (3) \emph{bug discovery time} (triggering bugs faster \ttbToprSF and \ttbToprAflgo, respectively).
Furthermore, \topr's balance of speed \emph{and} precision enables it to find \NumBugs new bugs in \NumBuggyProjects open-source applications, with \NumConfirmedBugs confirmed by developers, \hdfBugsHigh bugs labelled as ``Priority - 1. High'', and \NumFixedBugs bugs fixed --- underscoring the effectiveness of our framework.
\end{abstract}


\section{Introduction}
\label{sec::intro}

Directed fuzzing is a dynamic testing technique well-suited to testing scenarios like patch testing, bug reproduction, test case augmentation and regression testing \wrt specific targets in code~\cite{aflgo,cnid}.
Unlike conventional fuzzing, which aims to explore a program in its entirety, directed fuzzing instead focuses testing only on a \emph{pre-targeted} set of program locations, e.g., the source code line(s) affected by a patch.
To date, a variety of directed fuzzing techniques have emerged, including AFLGo~\cite{aflgo}, Beacon~\cite{beacon}, Hawkeye~\cite{chen_hawkeye_2018}, and SieveFuzz~\cite{sievefuzz}.

Like general-purpose fuzzers~\cite{fioraldi_afl_2020,serebryany_continuous_2016}, directed fuzzers must be both \emph{fast} and \emph{precise}.
Maintaining high speed is critical to finding software defects quickly (e.g., for timely patch fixing)~\cite{nagy_same-coverage_2021}, while precision is key to avoiding false positives and negatives (e.g., missed or spurious bugs)~\cite{winnie_2020}.
While many advancements have made general-purpose fuzzing fast~\cite{nagy_full-speed_2019, nagy_same-coverage_2021, winnie_2020}, most directed fuzzers unfortunately remain far slower.
The most popular approach for ``directing'' fuzzing to specific target locations is \emph{distance minimization}~\cite{aflgoref}, which iteratively selects test cases with the smallest distances relative to the target sites.
Yet, the runtime expense of distance minimization---extra program instrumentation to measure and track distances \emph{per} test case---is shown to deteriorate fuzzing speed by upwards of 92\%~\cite{sievefuzz}.

To overcome the performance bottleneck of distance minimization, recent directed fuzzers are instead adopting \emph{code~pruning} (e.g., SieveFuzz~\cite{sievefuzz}, ParmeSan~\cite{osterlund_parmesan_2020}).
At a high level, code pruning performs static control-flow analysis to identify and eliminate program paths deemed \emph{irrelevant} to reaching a given target location.
By preventing execution from being wasted on fruitless paths, directed fuzzers are shown to achieve upwards of 100\% higher test case throughput~\cite{sievefuzz}.

Unfortunately, pruning-based directed fuzzers are limited in \emph{precision}.
Current approaches universally rely on static analysis to generate control-flow graphs and enumerate relevant paths. 
Yet, recovering \emph{indirect} control flow statically is infeasible, preventing these fuzzers from exploring many critical paths (e.g., indirect call targets).
To improve precision, static analysis is often supplemented with additional \emph{dynamic} analysis to reanalyze control flow. 
However, the frequency of indirect edges (often tens of thousands per program) demands constant control-flow reanalysis, diminishing the otherwise high speeds of pruning-based directed fuzzers.
Thus, achieving a high directed fuzzing performance---without sacrificing precision---remains a compelling challenge.

This paper introduces the idea of \emph{enhanced code pruning}---a layering of conventional static-based code pruning with lightweight analysis heuristics to recover the overwhelming majority of indirect control flow: indirect function calls.
We demonstrate how enhanced code pruning facilitates \emph{fully static} control-flow analysis for pruning-based directed fuzzing, forgoing the high runtime overhead faced by conventional static/dynamic hybrid approaches~\cite{sievefuzz, osterlund_parmesan_2020}.
We implement our approach as a directed fuzzing prototype, \topr (\textbf{T}arget \textbf{O}riented \textbf{Pr}uning), and evaluate it alongside the leading pruning-based and pruning-agnostic directed fuzzers SieveFuzz~\cite{sievefuzz} and AFLGo~\cite{aflgo}.

We show that \topr's fast \emph{and} precise pruning achieves superior directed fuzzing effectiveness---and speed---across \textbf{24} real-world software benchmarks:
on average, \topr achieves \textbf{\tctToprSF} and \textbf{\tctToprAflgo} higher test case throughput and achieves \textbf{\trcToprSF} and \textbf{\trcToprAflgo} more target-relevant coverage than SieveFuzz and AFLGo, respectively.
We further demonstrate how \topr's efficient indirect edge recovery enables more expeditious directed fuzzing: on average, \topr is the fastest to discover 
bugs---outperforming SieveFuzz and AFLGo by \textbf{\ttbToprSF} and \textbf{\ttbToprAflgo}, respectively.
In an additional bug-finding case study on the latest versions of \textbf{9} real-world programs, \topr discovers \NumBugs new bugs---with \NumConfirmedBugs confirmed and \NumFixedBugs fixed by developers and \hdfBugsHigh bugs labelled as ``Priority - 1. High''.

In summary, this paper makes the following contributions:
\begin{itemize}
    \item We detail the challenges of pruning-based directed fuzzing. 
    We show that existing code pruning approaches are fundamentally unable to balance performance with precision, leading to cascading effects that impede directed fuzzing effectiveness.
    \item We introduce the idea of \emph{enhanced code pruning}: a technique by which conventional static analysis is augmented with additional, lightweight heuristics to capture the vast majority of problematic indirect edges.
    \item We implement enhanced code pruning as a prototype directed fuzzer, \topr, and evaluate it alongside the leading pruning-based directed fuzzer SieveFuzz~\cite{sievefuzz} and pruning-agnostic directed fuzzer AFLGo~\cite{aflgo}.
    \item We evaluate \topr on \textbf{24} real-world software benchmarks, and show that its enhanced code pruning enables faster, more effective directed fuzzing.
\end{itemize}

\section{Background \& Motivation}
\label{sec::motiv}

Below, we briefly introduce the high-level concepts behind fuzzing and directed fuzzing.
We discuss the challenges of upholding performance \emph{and} precision in directed fuzzing, and weigh current directed fuzzers' fundamental trade-offs.

\subsection{Fuzzing}
Fuzzing (short for ``fuzz testing'') represents one of the most popular and successful dynamic testing approaches in use today.
At a high level, fuzzing operates by:
\begin{enumerate}
    \item Mutating test cases at random~\cite{zalewski_american_2017} or via well-defined input dictionaries~\cite{pham_smart_2019} or grammars~\cite{aschermann_nautilus_2019,blazytko_grimoire_2019}; 
    \item Executing each test case on the program under test; and 
    \item Preserving---and subsequently re-mutating---only those test cases that trigger interesting behavior (e.g., new code coverage or crashes), while discarding all others.
\end{enumerate}

A variety of tools and techniques have emerged over the last three decades advancing many of fuzzing's core aspects: increasing speed~\cite{nagy_full-speed_2019}, enhancing test case generation~\cite{aschermann_redqueen_2018} and mutation~\cite{lv_mopt_2019}, and streamlining analysis of fuzzer-discovered bugs~\cite{yan_exploitmeter_2017}. 
As fuzzing continues to gain recognition for discovering numerous critical program defects, many efforts are exploring tailoring fuzzing to specific use cases: emergent classes of bugs~\cite{nguyen_binary-level_2020}, and new software, environments~\cite{winnie_2020}, and architectures~\cite{nagy_breaking_2021}.
More recently, a fuzzing technique called \emph{directed fuzzing} substitutes conventional fuzzing's whole-program exploration with one targeting specific program regions: patched code~\cite{aflgo}, reported crashing lines~\cite{sievefuzz}, and regressions~\cite{regressiongreyboxfuzzing}.

\begin{table}[h!]
    \centering
    \renewcommand*{\arraystretch}{1.3}%
    
    \begin{tabular}[b]{ p{1.8cm} | p{0.9cm} | p{1.4cm} | p{1.2cm}  p{1.2cm} }
        \hline
            \multirow{2}{*}{\textbf{Tool}} &
            \multirow{2}{*}{\textbf{Year}} &
            \multirow{2}{*}{\textbf{Directedness}} &
            \multicolumn{2}{c}{{{\textbf{Indirect Control Flow}}}} \\
            
            & & &
            \textbf{Recovery} &
            \textbf{Correctness} \\
            
            
        \hline
        
        \color{black}{AFLGo~\cite{aflgoref}}
            & \color{black}{2017}
            & \color{red}{Minimization}
            & \color{red}{None}
            & \color{red}{\xmark}
        \\

        \color{black}{Hawkeye~\cite{chen_hawkeye_2018}}
            & \color{black}{2018}
            & \color{red}{Minimization}
            & \color{green}{Static}
            & \color{red}{\xmark}
	\\

        \color{black}{ParmeSan~\cite{osterlund_parmesan_2020}}
            & \color{black}{2020}
            & \color{red}{Minimzation}
            & \color{red}{Dynamic}
            & \color{green}{\cmark}
	\\

        \color{black}{WindRanger~\cite{windranger}}
            & \color{black}{2022}
            & \color{green}{Pruning}
            & \color{green}{Static}
            & \color{red}{\xmark}
	\\

        \color{black}{RLTG~\cite{rltg}}
            & \color{black}{2023}
            & \color{green}{Pruning}
            & \color{red}{Dynamic}
            & \color{green}{\cmark}
	\\

        \color{black}{SieveFuzz~\cite{sievefuzz}}
            & \color{black}{2023}
            & \color{green}{Pruning}
            & \color{red}{Dynamic}
            & \color{green}{\cmark}
	\\

        \hline

        \rowcolor{gray!10}
        \color{black}{\topr}
            & \color{black}{2023}
            & \color{green}{Pruning}
            & \color{green}{Static}
            & \color{green}{\cmark}
        \\

        \hline

    \end{tabular}
    \caption{Popular directed fuzzers and their design trade-offs. }
    \label{tab:comparison}
\end{table}

\subsection{Directed Fuzzing}
Directed fuzzing differs from conventional fuzzing in that it focuses testing only on \emph{specific} program locations.
Generally, locations to be targeted are determined either manually (e.g., a line changed by a patch), or automatically via static analysis tools (e.g., a location believed to contain a bug).
In the following, we compare and contrast the key design decisions of directed fuzzers: how to achieve \emph{directedness} and how to \emph{recover} control flow from the program under test.

\subsubsection{Achieving Directedness}\label{sec:background:directedness}
Unlike general-purpose fuzzers (e.g., AFL~\cite{zalewski_american_2017}, libFuzzer~\cite{serebryany_oss-fuzz_2017}) that aim to fully explore the target program as a whole, directed fuzzers instead focus on testing a specific set of program locations.
Existing directed fuzzers are generally implemented atop undirected ones, but incorporate a mechanism for \emph{directedness}: steering exploration to the pre-determined target location(s).
Historically, many directed fuzzers achieved directedness via \emph{distance minimization}~\cite{aflgo}.

At a high level, minimization operates by first constructing a gradient of the program with respect to all pre-determined target locations. 
For all generated test cases, distances are computed at the basic-block-level relative to the target sites.
Lastly, the fuzzer prioritizes those test cases observed to have smaller and smaller relative distances---leveraging this as a proxy for steering testing toward the intended locations.
Yet while distance minimization intuitively models program exploration as an optimization problem, it unfortunately cannot prevent exploration of execution paths that are \emph{irrelevant} to reaching target sites.
In such cases, minimization-directed fuzzers risk getting stuck in the gradient's \emph{local} minima, forcing the fuzzer to randomly select other test cases in hopes of finding the correct path to the \emph{global} minimum~\cite{aflgo,chen_hawkeye_2018}.

To overcome minimization's wasteful exploration, emerging directed fuzzers are adopting \emph{path pruning}~\cite{sievefuzz, osterlund_parmesan_2020}: selectively restricting program control flow from being explored based on its relevancy to the intended target location(s).
Typically, pruning performs analysis to recover all program paths, generally via backwards-slicing from the intended region of interest (e.g., a reported buggy line of code). 
Path restriction is achieved by either discarding non-reaching test cases \emph{before} they are executed; or \emph{during} execution by terminating once a non-reaching path is observed taken.
In most implementations, pruning is coupled with minimization to further steer testing toward the pre-targeted program locations~\cite{osterlund_parmesan_2020}. 

\subsubsection{Control Flow Recovery}

To correctly guide testing toward target locations, directed fuzzers must precisely recover the program's execution paths.
Existing approaches universally rely on \emph{static} analysis to recover program control flow. 
However, statically recovering \emph{indirect} branches (e.g., indirect call targets) remains undecidable in practice, leaving purely-static approaches unsound---missing or over-approximating the set of outgoing edges from an indirect call site.
To overcome this challenge, a number of directed fuzzers are employing \emph{hybrid} analysis: performing a best-guess static analysis (e.g., points-to alias anlaysis~\cite{chen_hawkeye_2018}), followed by dynamic analysis to incorporate and re-analyze control flow for any concrete indirect edges seen during fuzzing's many test case executions.

\subsection{Weaknesses of Existing Directed Fuzzers}

\begin{figure}
  \begin{center}
    \framebox{\includegraphics[width=0.95\linewidth]{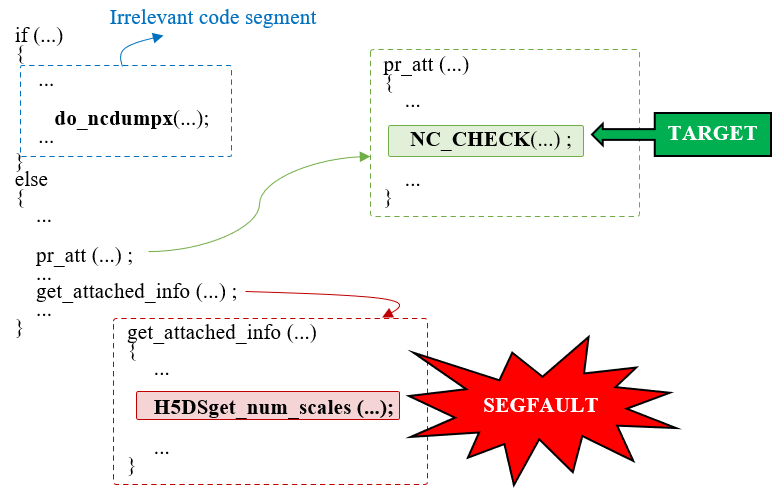}}
  \end{center}
  \caption{Simplified version of a memory corruption found by \topr but \emph{not}\Space{ found} by \aflgo in \texttt{NetCDF-C} v4.9.1 (latest version).}
  \label{fig:motivation:aflgo}
\end{figure}

In seeking to build an ideal directed fuzzer, we identify two key challenges that affect current directed fuzzers: (1) the high cost of directedness, and (2) the imprecision of indirect flow analysis.
We weigh existing directed fuzzers with respect to these challenges (shown in~\autoref{tab:comparison}); and elicit a criteria for a directed fuzzer to uphold both performance \emph{and} precision.

\subsubsection{The Cost of Directedness}
Directed fuzzers employing distance minimization suffer the risk of wasteful execution: spending time in execution paths that will \emph{not} reach the desired target locations.
This risk is compounded by the high cost of distance minimization's instrumentation, which adds significant runtime overhead per test case execution---shown to reduce test case throughput by upwards of 90\%~\cite{sievefuzz} relative to non-directed fuzzing.

\autoref{fig:motivation:aflgo} shows an example memory corruption bug in the latest version of the popular library \texttt{NetCDF-C}. 
We observe that \aflgo~\cite{aflgo}---the leading distance-minimization-based directed fuzzer---\emph{cannot} uncover this bug in 1~hour.
In examining the source code manually, we see a significant percentage of code paths that are not relevant to the target bug (e.g., the code containing call to function \texttt{do\_ncdumpx}). 
As \aflgo cannot filter-out such paths from being tested, much of its exploration will be spent exploring fruitless execution paths that will not reveal the program's bug.
Thus, we conclude that \textbf{path pruning achieves the best performance} as it prevents fuzzing from steering exploration down irrelevant paths.

\subsubsection{The Imprecision of Indirect Edge Recovery}

\begin{figure}
  \begin{center}
    \framebox{\includegraphics[width=0.95\linewidth]{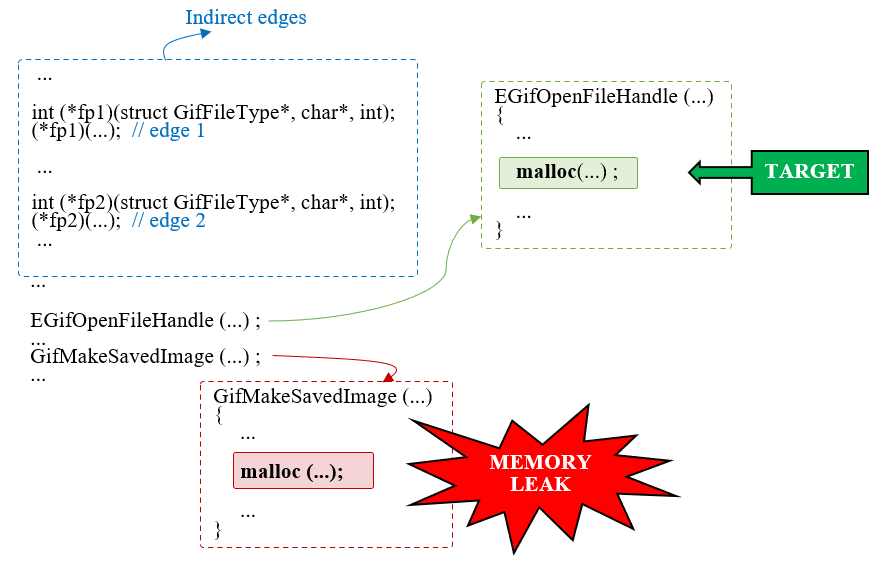}}
  \end{center}
  \caption{Simplified version of a memory leak found by \topr but \emph{not}\Space{ found} by SieveFuzz in \texttt{giflib} v5.2.1 (latest version).}
  \label{fig:motivation:sievefuzz}
\end{figure}

While most fuzzers aim to maximize exploration of the entire program, the goal of directed fuzzing is to instead focus only on \emph{specific} program regions (e.g., the lines affected by a program patch).
To this end, accurately modeling control flow is critical to knowing which paths to explore and which others to ignore.
Hawkeye~\cite{chen_hawkeye_2018} and WindRanger~\cite{windranger} attempt to resolve indirect control flow statically, which yields better performance as it requires just a one-time, pre-fuzzing analysis cost.
However, we observe that these tools rely on imprecise static analysis frameworks (e.g., SVF~\cite{sui_svf_2016}) that produce a significant over-approximation of possible paths, making them \emph{unusable} for general-purpose directed fuzzing.

Other directed fuzzers instead resolve indirect edges dynamically (e.g., SieveFuzz~\cite{sievefuzz}, ParmeSan~\cite{osterlund_parmesan_2020}, and RLTG~\cite{rltg}), which yields higher precision as their reachability analysis is expanded to include concrete edges. However, the reachability is dependent on the set of seed inputs.
Hence, we see that even these approaches miss critical paths that are not covered by the seed inputs.
\autoref{fig:motivation:sievefuzz} shows a memory leak bug in the latest version of \texttt{giflib} missed by the leading pruning-directed fuzzer SieveFuzz.
We observe that hybrid directed fuzzers will only incorporate indirect edges \emph{covered} by set of seed inputs, leaving paths that contain uncovered indirect edges missing.
Moreover, as edges are observed, these tools' reachability analyses need to be re-performed---deteriorating their performance further on programs where indirect edges are common.

In weighing the trade-offs of handling indirect edges, we conclude that the least-invasive option with respect to performance is \textbf{performing recovery \textbf{statically}}.
However, \textbf{achieving precision demands that \emph{no} indirect edges are missed}.

\begin{tcolorbox}
\textbf{Motivation}: Attaining fast \emph{and} precise directed fuzzing demands (\textbf{1}) \textbf{\emph{pruning-based}} directedness and (\textbf{2}) \textbf{\emph{fully-static}} indirect edge recovery. 
To date, \textbf{we see no existing directed fuzzer meeting these criteria}.
\end{tcolorbox}


\subsection{Attempt at Overcoming these Limitations}

Our tool, \topr, attempts to address the above mentioned weaknesses of existing directed fuzzers:
\subsubsection{The Cost of Directedness}
\topr identifies and prunes off all irrelevant code segments \wrt target (\eg a code segment marked in blue in \autoref{fig:motivation:aflgo}), including entire functions (\eg \texttt{do\_ncdumpx}). As a result, \topr was able to effectively explore the program space after the target location and uncover the memory corruption bug shown in \autoref{fig:motivation:aflgo} in 1 hour. On the other hand, the leading distance-minimization-based directed fuzzer, \aflgo was not able to find this bug in 1 hour due to its expensive exploration of irrelevant code segments \wrt target.
\subsubsection{The Imprecision of Indirect Edge Recovery}
\topr's lightweight static analysis uses function signature matching to resolve indirect edges. As a result, \topr recovered the 2 indirect edges marked in blue in \autoref{fig:motivation:sievefuzz} on the paths to the target location. \topr was thus able to reach the target location and uncover the memory leak bug shown in \autoref{fig:motivation:sievefuzz} in 1 hour. Whereas, the leading pruning-based directed fuzzer, \sfuzz was unable to reach the target and consequently did not find the memory leak bug in 1 hour. This is due to the fact that \sfuzz resolves indirect edges dynamically by completely relying on the set of seed inputs. On examining the logs generated by \sfuzz, we learned that it failed to dynamically recover both of these indirect edges.

\section{\topr: Enhancing Static Path Pruning}

\begin{figure}
  \begin{center}
    \includegraphics[width=\linewidth]{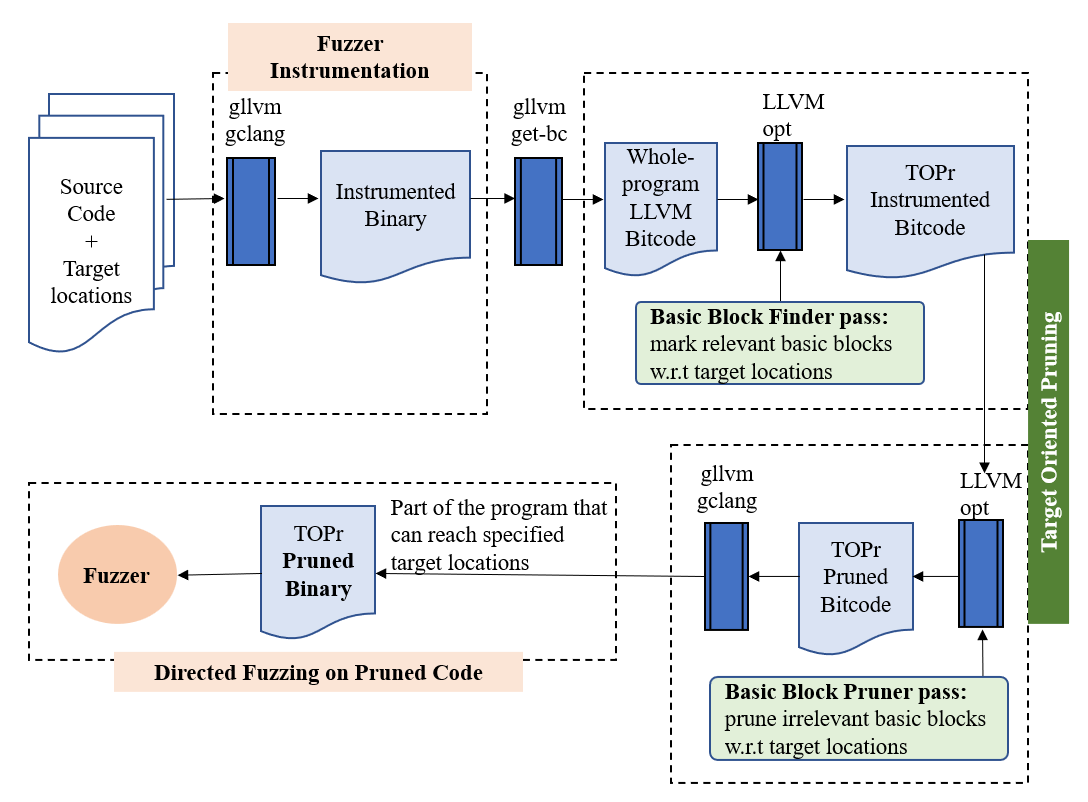}
  \end{center}
  \caption{High-level overview of our enhanced static code pruning for directed fuzzing.}
  \label{fig::topr::sys_overview}
\end{figure}

\autoref{fig::topr::sys_overview} details our high-level approach.
We perform backward slicing to isolate target locations---and their reaching paths---in the program. 
Our approach leverages a control-flow-based static analysis to identify basic blocks that can reach the target locations at the basic block level.
Following this analysis, we prune all irrelevant basic blocks, and perform directed fuzzing on the pruned program slice. 
We discuss the rest of our components and workflow below.

\subsection{Finding Relevant Control Flow}
\label{sec::bblfinder}

Our first phase identifies relevant basic blocks in program paths leading to desired target locations. 
Given a set of target locations, we perform a whole-program analysis as shown in \autoref{algo::pass1}.
Our approach leverages a conventional reachability analysis at the call and control-flow graphs to recover all basic blocks that reach target sites.
Below we detail our efforts to recover all direct and indirect control flow in a \emph{without} resorting to hybrid static/dynamic analysis.

\subsubsection{Recovering Direct Edges}
Using the default analyses provided by compilers (e.g., LLVM), we obtain all direct transfers at both the intra-  (e.g., \texttt{jmp .label}) and inter-procedural level  (e.g., \texttt{call foo}).
As this analysis is sound as-is, we need not incorporate additional heuristics to further strengthen its precision.

\subsubsection{Statically Recovering Indirect Edges}
While off-the-shelf control-flow analyses capture all direct edges, they are insufficient for recovering the frequent indirect edges ubiquitous among today's complex, real-world software. 
To overcome this challenge, we perform \emph{function signature matching} to handle indirect calls to functions in the whole-program module. 
We build on the basic approach in type-based pointer analysis and call-graph construction (e.g.,~\cite{LuHu19IndirectCalls, MilanovaETAL04CallGraphs}).

In case of call instructions that contain function pointers instead of names, we use signature matching to identify all functions in the whole-program module whose function signature matches with the function signature seen in the call instruction.
\texttt{getMatchedSignFnNames} on line 29 of algorithm \autoref{algo::pass1} accomplishes this task. 
This is a conservative mechanism which ensures that no relevant part of the program gets pruned out.
We construct the call graph for the entire program module and store it as a map whose key is a function and value is the set of functions that call the said function (\ie caller functions). This is a graph whose nodes are functions and there is an edge from function \texttt{f1} to function \texttt{f2} if \texttt{f2} can call \texttt{f1}. 
Effectively, it stores an inverse of the traditional call graph. 
Moreover, to identify call sites more precisely, we store function call sites (\ie call instructions) along with function names as nodes instead of having only functions.

For each target, we traverse the call graph bottom-up beginning from its parent function to identify all possible predecessor functions up to the program entry point.
For each function in this set, starting from the basic block of the target instruction or a function call instruction, we perform an iterative search to identify the relevant basic blocks (shown using \texttt{inverseDFS} on line 16).
All relevant blocks are marked using a call to a custom marker function (which we call the \emph{marker instruction}).
These marker instructions are inserted after the function call instructions that can reach targets (line 20).
Apart from these, all blocks in function calls appearing \emph{before} the target in relevant basic blocks are also marked as relevant (lines 29 and 37). 
Finally, we output the instrumented bitcode containing all basic block markers.

\begin{algorithm}
	\caption{Basic Block Finder Pass}\label{algo::pass1}
	\begin{algorithmic}[1]
        \STATE \textbf{Input}: Whole-program LLVM bitcode
        \STATE \textbf{Output}: \Space{Instrumented }LLVM bitcode with basic block markers
        \STATE requiredFns $\leftarrow$ $\emptyset$
	    \STATE targetInstructionSet $\leftarrow$ findAllTargets()
        \FOR {targetInstruction $\in$ targetInstructionSet}
    	\STATE targetFn $\leftarrow$ targetInstruction.getParentFn()
            \STATE targReachFnSet $\leftarrow$ \{targetFn\}
            \WHILE{targReachFnSet $\ne$ $\emptyset$}
                \STATE targReachFn $\leftarrow$ targReachFnSet.getNext()
    	      \STATE callerList $\leftarrow$ callGraph[targReachFn] 
    		    \WHILE{callerList $\ne$ $\emptyset$}
            	    \STATE callerFn $\leftarrow$ callerList.getNext()
                    \STATE targReachFnSet $\leftarrow$ targReachFnSet $\cup$ \{callerFn\}
                    \STATE callerFnIns $\leftarrow$ callerFn.ins
                    \STATE callerFnBbl $\leftarrow$ callerFnIns.getParentBbl()
                    \STATE predBblSet $\leftarrow$ inverseDFS(callerFnBbl)
                    \FOR {basicBlock $\in$ predBblSet}
                        \FOR {ins $\in$ basicBlock}
                            \IF {ins = callerFnIns}
                                \STATE insertMarkerFnCall()
                                \STATE \textbf{break}
                            \ELSE
                                \IF {ins.isCall()}
                                    \STATE fn $\leftarrow$ ins.getCalledFnName()
                                    \IF {fn $\ne$ ""}
                                        \STATE requiredFns $\leftarrow$ requiredFns $\cup$ \{fn\}
                                    \ELSE
                                        \STATE fnSign $\leftarrow$ ins.getCallSign()
                                        \STATE requiredFns $\leftarrow$ requiredFns $\cup$ \\ \ \ getMatchedSignFnNames(fnSign)
                                    \ENDIF
                                \ENDIF
                            \ENDIF
                        \ENDFOR
                    \ENDFOR
        	    \ENDWHILE
    		\ENDWHILE
        \STATE markFns(requiredFns)
      \ENDFOR
	\end{algorithmic}
\end{algorithm}

\subsection{Pruning Irrelevant Control Flow}
\label{sec::bblpruner}
With our marked bitcode of target-relevant control flow in hand, our second phase aims to prevent wasteful fuzzing by restricting all \emph{irrelevant} program paths.
We terminate quickly during the execution.
This pass outputs the pruned LLVM bitcode.
As shown in \autoref{algo::pass2}, we first identify locations of marker instructions in every basic block (lines 5--16);
if no marker instruction is found, then the entire basic block will be pruned (line 6); and if a marker instruction \emph{is} present as the last non-terminating instruction, then we skip the basic block (line 10).
We insert a call to \texttt{exit()} before all marker instructions (line 18). 
As \aflgo treats non-zero exit codes as erroneous program crashes, we intentionally use \texttt{exit(0)} to signal a normal termination.

\begin{algorithm}
	\caption{Basic Block Pruner Pass}\label{algo::pass2}
	\begin{algorithmic}[1]
        \STATE \textbf{Input}: \Space{Instrumented }LLVM bitcode with basic block markers
        \STATE \textbf{Output}: Pruned LLVM bitcode
	\FOR {fn $\in$ wholeProgram}
		\FOR {bbl $\in$ fn}
            \STATE toPrune $\leftarrow$ True
            \STATE pruneStartIns $\leftarrow$ bbl.begin()
            \FOR {ins $\in$ bbl}
                \IF {ins.isMarkerFnCall()}
                    \IF {ins.isLast()}
                        \STATE toPrune $\leftarrow$ False
                    \ELSE
                        \STATE pruneStartIns $\leftarrow$ ins
                    \STATE \textbf{break}
                    \ENDIF
                \ENDIF
            \ENDFOR
            \IF {toPrune}
                \STATE pruneStartIns.insertExitCallBefore()
            \ENDIF
        \ENDFOR
	\ENDFOR
    \end{algorithmic}
\end{algorithm}

\section{Implementation}
\label{sec::impl}

Below we detail our implementation of enhanced static code pruning: \topr (\textbf{T}arget \textbf{O}riented \textbf{Pr}uning).

\subsection{Overview}

We use gllvm~\cite{gllvm}, an LLVM~\cite{llvm} framework based toolchain for static analysis and rewriting (\ie pruning) of code. 
gllvm is wrapper around LLVM based compilers that provides tools to build whole-program (or whole-library) LLVM bitcode files.
\texttt{gclang} generates a whole-program executable.
Next, the \texttt{get-bc} tool is used to extract LLVM bitcode from this binary. 
The modular LLVM optimizer and analyzer, \texttt{opt} is then used to analyze and transform the bitcode. 
We integrated \topr atop a recent version of the directed greybox fuzzer, \aflgo~\cite{aflgoref}, which is commit \texttt{b170fad} that uses LLVM version 11.0. 

\autoref{fig::topr::sys_overview} provides the workflow of our framework. 
The user specifies the target locations in the source code of the program. \aflgo takes as input the source code and these targets to generate an instrumented binary for fuzzing.
We wrap gllvm around the \aflgo compiler in order to extract the LLVM bitcode from whole-program (or whole-library) instrumented binary that it generates. 
This instrumented bitcode is then pruned by \topr before being used for fuzzing. 
We run our transformation passes for pruning using the LLVM optimizer, \texttt{opt}. 
The pruned bitcode is finally compiled to a pruned binary using \texttt{gclang} and then used to perform directed fuzzing.

\subsection{Backward Slicing}
Many of the advanced analyses that are part of the LLVM project, such as alias analyses and memory SSA, are only intra-procedural and therefore insufficient:
LLVMSlicer~\cite{llvmslicer} is obsolete, while others (e.g.,~\cite{dgslicer,semslice,paraslicer,sympas}) are limited in their capabilities.
As we are not aware of an existing, fully-functional LLVM pass that performs backward slicing, we manually implement backward slicing atop our LLVM analyses.

\subsection{Fuzzing Component}
\topr's enhanced static path pruning is implemented atop the \aflgo directed fuzzer.
We leverage the fuzzer's lightweight instrumentation to obtain code coverage information (e.g., control-flow edges and hit count frequencies), alongside target-relative distances, for every generated input.

\section{Evaluation}
\label{sec::eval}

\newcommand{\rqbug}{\textbf{Research Question 1}}
\newcommand{\rqttb}{\textbf{Research Question 2}}
\newcommand{\rqtreach}{\textbf{Research Question 3}}
\newcommand{\rqthrp}{\textbf{Research Question 4}}

We evaluate \topr against the following two tools:
\begin{enumerate}
\item the leading distance-minimization-based directed fuzzer, \aflgo~\cite{aflgoref}.
\item the state-of-the-art pruning-based directed fuzzer SieveFuzz~\cite{sievefuzz}.
\end{enumerate}
Our evaluation addresses the following research questions:
\begin{enumerate}
\item \rqbug\xspace. How many unique bugs are found by the tools?
\item \rqttb\xspace. What is the time taken by the tools to expose the bugs?
\item \rqtreach\xspace. How well do the tools perform \wrt target reachability?
\item \rqthrp\xspace How fast are the tools in terms the test case throughput?
\end{enumerate}

\subsection{Evaluation Projects}
\autoref{tab::topr-eval-projs-patch} and \autoref{tab::topr-eval-projs-bug} provide details on the 24 real world benchmarks used in evaluation.
Based on the use case for evaluation, the benchmarks are split into two sets --- \ci Patch testing where the changed lines of code are set as target locations and \cii Bug reproduction where previously found bugs are set as target locations.
Moreover, based on the project versions, these benchmarks are categorized into two sets: 
\begin{enumerate}
    \item \textbf{Older project versions evaluated by \aflgo and \sfuzz.}
    This set consists of 15 of the 24 benchmarks. The benchmark IDs with suffixes \texttt{-2,-3,-4} fall into this category.
    These have been used previously in the evaluation of \aflgo~\cite{aflgoref,aflgorepoprojs} and \sfuzz~\cite{sievefuzz}.
    Most of these projects have been updated since they were evaluated by \aflgo and \sfuzz. Hence, finding bugs in these older versions has no benefit for the current code versions because the bugs had been already discovered and mostly fixed. However, these benchmarks are still useful for comparing \topr with these tools.
    \item \textbf{Latest project versions.} This set consists of the remaining 9 of the 24 benchmarks. The benchmark IDs with suffix \texttt{-1} fall into this category.
    It includes the latest versions of 7 projects from the first set, \ie all projects that changed since prior evaluations (the \texttt{libming} library has been abandoned for years, motivating the need to compare techniques on the latest project versions rather than old versions with known prior bugs). Furthermore, this set also includes the latest versions of 2 additional projects---NetCDF-C and HDF5. These are libraries are extensively used to analyze and process scientific datasets in geo-science applications~\cite{hdf5,netcdf}.
    Finding new bugs in the latest versions of these projects provides more confidence in assessing and establishing the significance of \topr.
\end{enumerate}

\begin{table}[!h]
    \centering
    \renewcommand*{\arraystretch}{1.3}%

    \caption{\emph{Patch testing} benchmarks (all latest versions except libxml2-2) used in our evaluation.}

    
    \label{tab::topr-eval-projs-patch}

\begin{tabular}{ l|c }
    \specialrule{.1em}{0em}{0em} 
    \hline
        \mystrut
        \textbf{Benchmark} & \textbf{Commit ID} \\ 
    \hline

    \rowcolor{gray!10}cxxfilt-1 
        & 32778522c7d 
        \\
    giflib-1 
        & adf5a1a 
        \\
    \rowcolor{gray!10}hdf5 
        & 0553fb7 
        \\
    jasper-1 
        & 402d096 
        \\
    \rowcolor{gray!10}libxml2-1 
        & f507d167 
        \\
    libxml2-2 
        & ef709ce2 
        \\
    \rowcolor{gray!10}lrzip-1 
        & e5e9a61 
        \\
    mjs-1 
        & b1b6eac 
        \\
    \rowcolor{gray!10}netcdf-c 
        & 63150df 
        \\
    objdump-1 
        & 32778522c7d
        \\
        
    \hline
\end{tabular}
\end{table}
\begin{table}[!h]
    \centering
    \renewcommand*{\arraystretch}{1.3}%

    \caption{\emph{Bug reproduction} benchmarks in our evaluation.}

    
    \label{tab::topr-eval-projs-bug}

\begin{tabular}{ l|c|c|c }
    \specialrule{.1em}{0em}{0em} 
    \hline
        \mystrut
        \textbf{Benchmark} & \textbf{Commit ID} & \textbf{Error ID} & \textbf{Error Type} \\
    \hline

    \rowcolor{gray!10}cxxfilt-2 
        & 2c49145
        & CVE-2016-4487 
        & Heap UAF \\

    giflib-2 
        & 72e31ff 
        & Bug \#74 
        & Double Free \\
    \rowcolor{gray!10}giflib-3 
        & adf5a1a
        & N/A
        & Memory Leak \\
        
    jasper-2 
        & 142245b
        & CVE-2015-5221 
        & Heap UAF \\
    \rowcolor{gray!10}jasper-3 
        & 142245b
        & N/A
        & Memory Leak \\
        
    libming-1 
        & b72cc2f
        & CVE-2018-8807 
        & Heap UAF \\
    \rowcolor{gray!10}libming-2 
        & b72cc2f 
        & CVE-2018-8962 
        & Heap UAF \\
    libming-3 
        & b72cc2f
        & N/A
        & Memory Leak \\
        
    \rowcolor{gray!10}lrzip-2 
        & ed51e14 
        & CVE-2018-11496 
        & Heap UAF \\
    lrzip-3 
        & 9de7ccb 
        & CVE-2017-8846 
        & Heap UAF \\
        
    \rowcolor{gray!10}mjs-2 
        & d6c06a6
        & Issue \#57 
        & Int Overflow \\
    mjs-3 
        & 9eae0e6
        & Issue \#78 
        & Heap UAF \\
    \rowcolor{gray!10}mjs-4 
        & 2827bd0 
        & N/A
        & Floating Point \\
        
    objdump-2 
        & a6c21d4a553
        & CVE-2017-8392 
        & Invalid Read \\

    \hline
\end{tabular}
\end{table}

\subsection{Metrics}
The evaluation metrics measure target specific statistics to effectively compare the tools against one another.
These metrics are obtained by replaying the inputs generated during fuzzing on code instrumented using LLVM to compute these metrics. Note that all the metrics that we use in our evaluation are \emph{target specific}, \ie all these metrics are computed only with reference to program paths and basic blocks that reach target locations.
\autoref{tab::metrics} summarizes the following metrics used in our evaluation: 
\begin{itemize}
    \item \textbf{Unique Bugs and Unique Traces.} We count only the bugs on program paths that include target locations. Note that the program path for a given input may not reach the target location even for \topr that guides exploration towards targets. First, the program may crash before a target is reached, \eg \segv. Such an input/path would not show any bug in the target itself but shows that some paths to the target have bugs. Second, the program path may follow a branch that cannot reach a target, \eg the target may be in the \emph{else} block of an \emph{if} statement, while the input follows the \emph{then} block. (The key contribution of \topr is indeed to terminate such paths as soon as they reach a basic block from which the target cannot be reached.)
    We use the ASAN tool to detect bugs during replay of inputs generated by the fuzzer. Of the bugs reported by ASAN, we filter out and deduplicate ASAN reports to count --- \ci Unique bugs: based on \emph{primary bug location} that accounts for only the primary source file name and line number where the bug was triggered and \cii Unique traces: based on \emph{full stack trace} that accounts for all source file names and line numbers listed in the error stack trace.
    The number of unique bugs [\rqbug] is the most important metric used to evaluate the efficacy of fuzzing in our evaluation and thus to compare \topr against \aflgo and \sfuzz.
     \item \textbf{Time to Bug Exposure.} This is a measure of the time taken by the fuzzer to generate the input that triggers the first bug in the program [\rqttb].
    \item \textbf{Target Reachability.} We measure target-relevant coverage in terms of basic blocks covered in the pruned code. 
    The other metrics include the number of times that one or more targets are reached and the number of inputs that reach at least one of the target locations during execution. 
    These metrics put together conceptually show how much the generated inputs ``explore'' the targets [\rqtreach]. A tool that reaches the target more often is likely to have a higher chance to detect bugs related to targets.
    \item \textbf{Test Case Throughput.} This is a measure of the total number of executions during the fuzzing campaign used to evaluate the speed of fuzzing [\rqthrp].
\end{itemize}

\begin{table}[!h]
    \centering
    \renewcommand*{\arraystretch}{1.3}%

    \caption{Metrics used in evaluation.}

    
    \label{tab::metrics}

\begin{tabular}{ l|l }
    \specialrule{.1em}{0em}{0em} 
    \hline
        \mystrut
        \textbf{No.} & \textbf{Metric} \\ 
    \hline

    \rowcolor{gray!10}1
        & Unique bugs based on primary location in ASAN trace
        \\
     2
        & Unique traces based on full ASAN trace
        \\
     \rowcolor{gray!10}3
        & Time taken to exposure of first bug within fuzzing campaign
        \\
     4
        & Target-Relevant coverage in terms of basic blocks
        \\
     \rowcolor{gray!10}5
        & Number of target reaches
        \\
    6
        & Number of target reaching inputs
        \\
    \rowcolor{gray!10}7
        & Test case throughput
        \\
    \hline
\end{tabular}
\end{table}

\subsection{Experimental Setup}
We run experiments on a Virtual Machine with Ubuntu 20.04 OS, 66GB Memory, Intel® Xeon(R) CPU E5-2620 @ 2.00GHz × 24.  The time-to-exploitation is set to 45 minutes and the timeout is set to 1 hour for the fuzzing campaigns. The rationale for setting a 1 hour time limit is as follows.
Most use-case scenarios for targeted fuzzing require relatively \emph{faster} feedback than traditional fuzzing that explores an entire program.
For example, consider the use scenario of regression testing~\cite{regtest}---validating whether a recent code commit in a large code base breaks some functionality~\cite{dise}.
While traditional fuzzing may explore the entire large code base to search for potential bugs, targeted fuzzing would focus only on the changed code to explore whether it itself introduced a bug.
In others words, generating a test input that leads to a failure in the unchanged code can be considered a kind of ``false alarm'' because the developer may be already aware of those existing bugs.
For this reason, targeted fuzzing has a much smaller time budget than traditional fuzzing (which in the limit can run continuously ``forever'' and in research experiments is typically evaluated with 24-hour-long experiments).
A reasonable expectation is that targeted fuzzing finishes in the same time it takes to run the regression test suite, e.g., with targeted fuzzing running in parallel with the traditional regression test suite.
As a result, in our experiments we use 1-hour-long time limit for running targeted fuzzing.

To ensure a fair evaluation, we follow the experimental setup (to the extent possible) used in prior evaluation of the tools, \aflgo~\cite{aflgoref} and \sfuzz~\cite{sievefuzz}.
For evaluation of \topr against \aflgo, the fuzzing campaigns make use of 22 out of the 24 cores. 
For evaluation of \topr against \sfuzz, the fuzzing campaigns make use of 1 core and 10 trials of fuzzing campaigns are conducted. In this case, we use the arithmetic mean to report the summary of results across the 10 trials.

\subsection{Results}

\noindent \Note that $\color{red}{X}$ in the tables indicates that a tool failed to generate a value \wrt the given metric. \sfuzz failed to run on cxxfilt-2 benchmark and therefore all corresponding values are shown as $\color{red}{X}$ in the tables.\\

\subsubsection{Unique Bugs and Unique Traces}
Tables \autoref{tab::topr-aflgo-bugs} and \autoref{tab::topr-sf-bugs} show the results for the number of unique bugs found and the number of unique ASAN traces for \topr vs.\ \aflgo and \topr vs.\ \sfuzz respectively. Higher values indicate better performance.
For most benchmarks, \topr finds more bugs than its competitors, \aflgo and \sfuzz. For the remaining benchmarks, \topr performs on a par with the other 2 tools. This shows that in terms of the most important evaluation metric, unique bugs found, \topr outperforms both \aflgo and \sfuzz (highlighted by numbers shown in bold in the tables). 
\topr \emph{found \NumBugs bugs in the latest versions of \NumBuggyProjects projects. 
} More details on the reported bugs are discussed in Section~\autoref{sec::eval::bugs}.

\subsubsection{Time to Bug Exposure}
Tables \autoref{tab::topr-aflgo-ttb} and \autoref{tab::topr-sf-ttb} show the results for time to exposure of the first bug within the fuzzing campaigns along with type and location of the bugs. Lower values for TTB indicate better performance.
For most benchmarks, \topr finds bugs faster (lower TTB) in comparison to its competitors, \aflgo and \sfuzz. The average speedup of \topr\xspace \wrt time to bug exposure in Tables \autoref{tab::topr-aflgo-ttb} and \autoref{tab::topr-sf-ttb} are \textbf{\ttbToprAflgo} and \textbf{\ttbToprSF} respectively (We compute this average for all benchmarks where values for both tools are not $\color{red}{X}$, and we compute the average of $(TOPr-baseline)/baseline$ values where $baseline$ is \aflgo or \sfuzz respectively). This indicates that once again \topr outperforms both \aflgo and \sfuzz (highlighted by numbers shown in bold in the tables).

\subsubsection{Target Reachability and Test Case Throughput}
Tables \autoref{tab::topr-aflgo-other} and \autoref{tab::topr-sf-other} show the results for basic block coverage in target-relevant code, number of times target(s) reached, number of inputs reaching target(s) and the test case throughput. Higher values indicate better performance. For most of the benchmarks, \topr achieves better target rechability compared to its competitors, \aflgo and \sfuzz.
On average, \topr's increase in test-case throughput in Tables \autoref{tab::topr-aflgo-other} and \autoref{tab::topr-sf-other} are \textbf{\tctToprAflgo} and \textbf{\tctToprSF} respectively.
\topr's average increase in target relevant coverage as shown in Tables \autoref{tab::topr-aflgo-other} and \autoref{tab::topr-sf-other} are \textbf{\trcToprAflgo} and \textbf{\trcToprSF} respectively.
Yet again, \wrt target reachability, \topr outperforms both \aflgo and \sfuzz (highlighted by numbers shown in bold in the tables). \\

SieveFuzz operates by pruning irrelevant control-flow at a function-level granularity, and incorporates statically-unobtainable indirect edges in its dynamically-updated CFG. As shown in \autoref{tab::topr-sf-bugs}, \autoref{tab::topr-sf-ttb} and \autoref{tab::topr-sf-other}, \topr enables a more thorough exploration of directed fuzzing target locations across the 6 benchmarks---reaching target locations at a higher frequency per trial---enabling upwards of 12$\times$ more unique bugs to be discovered than SieveFuzz. We posit that \topr's improvement over SieveFuzz is primarily due to SieveFuzz's coarse-grained pruning coupled with its higher overhead: \topr's pruning instead operates at the basic block level and captures indirect function call edges statically, thus achieving higher precision \emph{and} speed than SieveFuzz.

\textbf{Summary.} To summarize the results overall, \topr outperforms both \aflgo and \sfuzz in terms of \ci bug discovery (both number and speed), \cii target reachability and \ciii speed by achieveing a higher test case throughput.

\renewcommand{\arraystretch}{1.2}
\begin{table}[!htbp]
    \centering
    \caption{\aflgo vs.\ \topr: number of AddressSanitizer-reported unique bugs found (\textit{UB}) and unique bug stack traces (\textit{UT}) for \topr versus \aflgo. Numbers are shown \textbf{in bold} when \topr outperforms \aflgo. (\emph{Higher is better.})}
    
    \label{tab::topr-aflgo-bugs}
  
\begin{tabular}{l|cc|cc}

    \specialrule{.1em}{0em}{0em} 
    \hline
        \mystrut
        
    \textbf{Benchmark} & \textbf{\aflgo$_{UB}$} & \textbf{\topr$_{UB}$} & \textbf{\aflgo$_{UT}$} &  \textbf{\topr$_{UT}$} \\

    \hline

\rowcolor{gray!10}cxxfilt-2 
    & 18 
    & \textbf{32} 
    & 134 
    & \textbf{310} \\
giflib-1 
    & 4 
    & 4 
    & 32
    & \textbf{46} \\
\rowcolor{gray!10}giflib-2 
    & 2 
    & 2 
    & 2 
    & 2 \\
hdf5 
    & 11 
    & \textbf{12} 
    & 16 
    & 16 \\
\rowcolor{gray!10}jasper-2 
    & 15 
    & \textbf{24} 
    & 271 
    & \textbf{433} \\

libming-1 
    & 0 
    & \textbf{67} 
    & 0 
    & \textbf{2126} \\

\rowcolor{gray!10}libming-2 
    & 0 
    & \textbf{77} 
    & 0 
    & \textbf{2101} \\
libxml2-2 
    & 8 
    & \textbf{9} 
    & 14 
    & \textbf{20} \\
\rowcolor{gray!10}lrzip-1 
    & 1 
    & 1 
    & 1 
    & 1 \\
lrzip-2 
    & 9 
    & 9 
    & 22
    & 22 \\
\rowcolor{gray!10}lrzip-3 
    & 1 
    & 1 
    & 1 
    & 1 \\
mjs-1 
    & 0
    & \textbf{1} 
    & 0 
    & \textbf{1} \\
\rowcolor{gray!10}mjs-2
    & 13 
    & \textbf{14} 
    & 22 
    & \textbf{23} \\
mjs-3 
    & 2 
    & 2 
    & 2 
    & 2 \\
\rowcolor{gray!10}netcdf-c
    & 2 
    & \textbf{6} 
    & 3 
    & \textbf{9} \\
objdump-2 
    & 2 
    & 2 
    & 2 
    & 2 \\
\hline

\end{tabular}
\end{table}

\renewcommand{\arraystretch}{1.2}
\begin{table}[!h]
    \scriptsize
    \centering

    \caption{\aflgo vs.\ \topr: time to bug exposure in seconds (TTB) of the \emph{first} bug within the fuzzing campaign. Numbers are shown \textbf{in bold} when \topr outperforms \aflgo. (\emph{Lower is better.})}
    
    \label{tab::topr-aflgo-ttb}
  
\begin{tabular}{p{1.15cm}|p{0.4cm}p{0.4cm}p{1.6cm}|p{0.4cm}p{0.4cm}p{1.6cm}}

    \specialrule{.1em}{0em}{0em} 
    \hline

    \multirow{2}{*}{\textbf{Benchmark}} &
        \multicolumn{3}{c|}{\textbf{\aflgo}} &
        \multicolumn{3}{c}{\textbf{\topr}} \\

    & \textbf{TTB} & \textbf{Type} & \textbf{Location} 
    & \textbf{TTB} & \textbf{Type} & \textbf{Location}\\
    & \textbf{(sec)} &  &  
    & \textbf{(sec)} &  & \\
    \hline

\rowcolor{gray!10}\multirow{2}{*}{} netcdf-c 
    & 21 & Leak & nchashmap.c:183 
    & \textbf{20} & Leak & nchashmap.c:183 \\

\multirow{2}{*}{} hdf5 
    & 51 & Leak & h5trav.c:124 
    & \textbf{35} & Leak & h5trav.c:124 \\

\rowcolor{gray!10}\multirow{2}{*}{} mjs-1 
    & \color{red}{X} & \color{red}{X} & \color{red}{X} & \textbf{510} 
    & Crash & Unspecified \\

\multirow{2}{*}{} lrzip-1 
    & 1873 & Leak & stream.c:1692 
    & 3443 & Heap & stream.c:1692 \\

\rowcolor{gray!10}\multirow{2}{*}{} giflib-1 
    & 357 & Leak & gifalloc.c:329 
    & \textbf{92} & Leak & egif\_lib.c:101 \\

\multirow{2}{*}{} libming-1 
    & \color{red}{X} & \color{red}{X} & \color{red}{X} & \textbf{506} 
    & Leak & parser.c:2435 \\

\rowcolor{gray!10}\multirow{2}{*}{} libming-2 
    & \color{red}{X} & \color{red}{X} & \color{red}{X} 
    & \textbf{2450} & Heap & decompile.c:896 \\

\multirow{2}{*}{} libxml2-2 
    & 241 & Heap & xmlsave.c:2057 
    & 1158 & Leak & valid.c:952 \\

\rowcolor{gray!10}\multirow{2}{*}{} jasper-2 
    & 31 & Leak & jas\_malloc.c:106 
    & \textbf{10} & Leak & jas\_malloc.c:106 \\

\multirow{2}{*}{} cxxfilt-2 
    & 256 & Heap & xmalloc.c:147 
    & \textbf{62} & Heap & xmalloc.c:147 \\

\rowcolor{gray!10}\multirow{2}{*}{} mjs-2 
    & 764 & Stack & mjs.c:12522 
    & \textbf{384} & Crash & mjs.c:8820 \\

\multirow{2}{*}{} mjs-3 
    & 1813 & Crash & mjs.c:9082 
    & 2721 & Crash & mjs.c:9082 \\

\rowcolor{gray!10}\multirow{2}{*}{} objdump-2 
    & 247 & Heap & xmalloc.c:147 
    & \textbf{67} & Heap & xmalloc.c:147 \\

\multirow{2}{*}{} giflib-2 
    & 41 & Crash & gifsponge.c:58 
    & 56 & Crash & gifsponge.c:58 \\

\rowcolor{gray!10}\multirow{2}{*}{} lrzip-2 
    & 15 & Leak & stream.c:1651
    & \textbf{5} & Leak & stream.c:1651 \\

\multirow{2}{*}{} lrzip-3 
    & 5 & Leak & lrzip.c:1326 
    & 5 & Leak & lrzip.c:1326 \\

\hline

\end{tabular}
\end{table}

\renewcommand{\arraystretch}{1.2}
\begin{table*}[!ht]
  \centering
  \caption{\aflgo vs.\ \topr: basic block coverage of target-relevant code, total instances target(s) is reached, total inputs reaching target(s), and test case throughput. Numbers are shown \textbf{in bold} when \topr outperforms \aflgo. (\emph{Higher is better.})}
  \label{tab::topr-aflgo-other}

\begin{tabular}{l|cc|cc|cc|cc}
    \specialrule{.1em}{0em}{0em} 
    \hline
        \mystrut
        
    \multirow{2}{*}{\textbf{Benchmark}} &
        \multicolumn{2}{c|}{\textbf{Target-Relevant Coverage}} &
        \multicolumn{2}{c|}{\textbf{Target Reaches}} & 
        \multicolumn{2}{c|}{\textbf{Target-Reaching Inputs}} & 
        \multicolumn{2}{c}{\textbf{Test Case Throughput}} \\
        & \aflgo & \topr & \aflgo & \topr & \aflgo & \topr & \aflgo & \topr \\
\hline

\rowcolor{gray!10}cxxfilt-1 
    & 4.86\% & \textbf{4.94\%} 
    & 2777115 & \textbf{2856414} 
    & 60997 & \textbf{63397} 
    & 39121411 & 29160761 \\
cxxfilt-2 
    & 6.31\% & \textbf{6.48\%} 
    & 44674525 & 5572273 
    & 87367 & \textbf{116256} 
    & 12424857 & \textbf{23310188} \\
\rowcolor{gray!10}giflib-1 
    & 65.19\% & \textbf{65.41\%} 
    & 34529 & \textbf{34563} 
    & 7075 & \textbf{7119} 
    & 31487585 & \textbf{31622273} \\
giflib-2 
    & 55.43\% & \textbf{57.74\%} 
    & 337 & 310 
    & 259 & 208 
    & 26010883 & \textbf{38766493} \\
\rowcolor{gray!10}hdf5 
    & 10.69\% & \textbf{10.8\%} 
    & 17430 & \textbf{23353} 
    & 17430  & \textbf{23353} 
    & 3817389 & \textbf{5857210} \\
jasper-1 
    & 31.29\% & \textbf{31.37\%} 
    & 1304630 & \textbf{3603540} 
    & 31622 & 25034 
    & 2220500 & \textbf{3064910} \\
\rowcolor{gray!10}jasper-2 
    & 38.39\% & \textbf{39.00\%} 
    & 43682 & \textbf{50691} 
    & 35066 & 34489 
    & 1451864 & \textbf{5378160} \\
libming-1 
    & 49.91\% & \textbf{65.03\%} 
    & 0 & \textbf{58533} 
    & 0 & \textbf{58533} 
    & 16480846 & 16244787 \\
\rowcolor{gray!10}libming-2 
    & 55.56\% & \textbf{67.75\%} 
    & 0 & \textbf{129321} 
    & 0 & \textbf{60198} 
    & 14670904 & \textbf{17524671} \\
libxml2-1 
    & 14.45\% & \textbf{15.42\%} 
    & 60107 & \textbf{106178} 
    & 54738 & \textbf{72002} 
    & 4354232 & \textbf{8956259} \\
\rowcolor{gray!10}libxml2-2 
    & 14.33\% & \textbf{15.31\%} 
    & 81886 & \textbf{142048} 
    & 66420 & \textbf{74602} 
    & 6717548 & \textbf{8976242} \\
lrzip-1 
    & 11.52\% & \textbf{15.86\%} 
    & 12250 & 1837 
    & 12250 & 1837 
    & 3679161 & \textbf{22457124} \\
\rowcolor{gray!10}lrzip-2 
    & 22.95\% & \textbf{25.39\%} 
    & 116250 & \textbf{188456} 
    & 12575 & \textbf{15132} 
    & 3476647  & \textbf{3732705} \\
lrzip-3 
    & 22.77\% & \textbf{24.22\%} 
    & 13924 & \textbf{16373} 
    & 13924 & \textbf{16373} 
    & 5139749 & 3715864 \\ 
\rowcolor{gray!10}mjs-1 
    & 39.21\% & \textbf{40.82\%} 
    & 155956 & \textbf{232189} 
    & 41547 & 38517 
    & 36098187 & 31482133 \\
mjs-2 
    & 38.75\% & \textbf{39.94\%} 
    & 2351387 & \textbf{2618489} 
    & 34629 & \textbf{43924} 
    & 23525141 & \textbf{36261339} \\
\rowcolor{gray!10}mjs-3 
    & 40.55\% & \textbf{41.13\%} 
    & 602653 & \textbf{690056} 
    & 36993 & \textbf{42194} 
    & 24022544 & \textbf{27349040} \\
netcdf-c 
    & 6.29\% & 6.26\% 
    & 4995 & \textbf{10815} 
    & 999 & \textbf{2163} 
    & 3419075 & \textbf{4730305} \\
\rowcolor{gray!10}objdump-1 
    & 2.89\% & \textbf{3.07\%}
    & 138228 & \textbf{338736} 
    & 64836 & 24029 
    & 9707018 & \textbf{14370966} \\
objdump-2 
    & 4.07\% & \textbf{5.11\%} 
    & 25975 & \textbf{29038} 
    & 23256 & 15947 
    & 2721294 & \textbf{8077430} \\
\hline
\end{tabular}
\end{table*}
\renewcommand{\arraystretch}{1}

\renewcommand{\arraystretch}{1.2}
\begin{table}[!htbp]
    \centering
    \caption{\sfuzz vs.\ \topr: mean number of AddressSanitizer-reported unique bugs found (\textit{UB}) and unique bug stack traces (\textit{UT}) for \topr versus \sfuzz. Numbers are shown \textbf{in bold} when \topr outperforms \sfuzz. (\emph{Higher is better.})}

    \label{tab::topr-sf-bugs}
    
\begin{tabular}{l|p{1.4cm}p{1.2cm}|p{1.4cm}p{1.2cm}}
    \specialrule{.1em}{0em}{0em} 
    \hline
        \mystrut
    \textbf{Benchmark} & \textbf{\sfuzz$_{UB}$} & \textbf{\topr$_{UB}$} & \textbf{\sfuzz$_{UT}$} &  \textbf{\topr$_{UT}$} \\
    \hline
\rowcolor{gray!10}\multirow{2}{*}{} cxxfilt-2 
    & \color{red}{X}
    & \textbf{2.6} 
    & \color{red}{X}
    & \textbf{6}  \\
\multirow{2}{*}{} giflib-3 
    & 0 
    & \textbf{0.8} 
    & 0 
    & \textbf{0.9} \\
\rowcolor{gray!10}\multirow{2}{*}{} jasper-3 
    & 1 
    & \textbf{10}
    & 3.8
    & \textbf{50} \\
\multirow{2}{*}{} libming-3 
    & 2.3
    & \textbf{3.1}
    & 2.5
    & \textbf{3.1} \\
\rowcolor{gray!10}\multirow{2}{*}{} lrzip-2 
    & 1 
    & \textbf{3.6} 
    & 1 
    & \textbf{7.8} \\
\multirow{2}{*}{} mjs-4 
    & 0.1
    & \textbf{1} 
    & 0.1 
    & \textbf{1} \\

\hline
\end{tabular}
\end{table}

\renewcommand{\arraystretch}{1.2}
\begin{table}[!h]
    \scriptsize
    \centering

    \caption{\sfuzz vs.\ \topr: mean time to bug exposure in seconds (TTB) of the \emph{first} bug within fuzzing campaign. Numbers are shown \textbf{in bold} when \topr outperforms \sfuzz. (\emph{Lower is better.})}
    
    \label{tab::topr-sf-ttb}
  
\begin{tabular}{p{1.05cm}|p{0.3cm}p{0.35cm}p{1.6cm}|p{0.45cm}p{0.35cm}p{2.05cm}}
    \specialrule{.1em}{0em}{0em} 
    \hline
    
    \multirow{2}{*}{\textbf{Benchmark}} &
        \multicolumn{3}{c|}{\textbf{\sfuzz}} &
        \multicolumn{3}{c}{\textbf{\topr}} \\

    & \textbf{TTB} & \textbf{Type} & \textbf{Location} 
    & \textbf{TTB} & \textbf{Type} & \textbf{Location}\\
    & \textbf{(sec)} &  &  
    & \textbf{(sec)} &  & \\
    \hline

\rowcolor{gray!10}mjs-4 
    & 3307 & FPE & mjs.c:8602
    & \textbf{2041.6} & Stack & mjs.c:12488 \\
    & & & & & FPE & mjs.c:8602 \\

giflib-3 
    & \color{red}{X} & \color{red}{X} & \color{red}{X}
    & \textbf{960.6} & Leak & gifalloc.c:329 \\

\rowcolor{gray!10}lrzip-2 
    & 7.7 & Leak & stream.c:1651
    & \textbf{4.8} & Leak & stream.c:1651 \\

cxxfilt-2 
    & \color{red}{X} & \color{red}{X} & \color{red}{X}
    & \textbf{1423} & Crash & cp-demangle.c:1596 \\
    & & & & & Crash & cplus-dem.c:4839 \\
    & & & & & Heap & xmalloc.c:147 \\
    & & & & & Leak & xmalloc.c:147 \\
    
\rowcolor{gray!10}jasper-3 
    & 5.4 & Crash & jas\_malloc.c:111
    & 31.4 & Crash & jas\_cm.c:1281 \\

libming-3 
    & 15.6 & Leak & read.c:227 
    & \textbf{5.4} & Leak & parser.c:882 \\

\hline

\end{tabular}
\end{table}

\renewcommand{\arraystretch}{1.2}
\begin{table*}[!ht]
  \centering
  \caption{\sfuzz vs.\ \topr: mean basic block coverage of target-relevant code, mean of instances target(s) reached, mean of inputs reaching target(s), and mean of test case throughput. Numbers are shown \textbf{in bold} when \topr outperforms \sfuzz. (\emph{Higher is better.})}
  \label{tab::topr-sf-other}

\begin{tabular}{l|cc|cc|cc|cc}
    \specialrule{.1em}{0em}{0em} 
    \hline
        \mystrut
        
    \multirow{2}{*}{\textbf{Benchmark}} &
        \multicolumn{2}{c|}{\textbf{Target-Relevant Coverage}} &
        \multicolumn{2}{c|}{\textbf{Target Reaches}} & 
        \multicolumn{2}{c|}{\textbf{Target-Reaching Inputs}} & 
        \multicolumn{2}{c}{\textbf{Test Case Throughput}} \\
        & \sfuzz & \topr & \sfuzz & \topr & \sfuzz & \topr & \sfuzz & \topr \\
\hline

\rowcolor{gray!10}cxxfilt-2 
    & \color{red}{X} & \textbf{4.72\%} 
    & \color{red}{X} & \textbf{2561} 
    & \color{red}{X} & \textbf{1054.6} 
    & \color{red}{X} & \textbf{1649432.1} \\

giflib-3 
    & 47.74\% & \textbf{52.33\%} 
    & 0 & \textbf{42.2} 
    & 0 & \textbf{42.2} 
    & 321822.4 & \textbf{1643788.5} \\

\rowcolor{gray!10}jasper-3 
    & 5.81\% & \textbf{40.64\%} 
    & 81.7 & \textbf{842.4} 
    & 81.7 & \textbf{842.4} 
    & 121633.9 & 111484.4 \\
    
libming-3 
    & 3.47\% & \textbf{3.63\%} 
    & 108.1 & 54 
    & 38.1 & 18 
    & 783136.5 & 211008.1 \\
    
\rowcolor{gray!10}lrzip-2 
    & 9.64\% & \textbf{20.68\%} 
    & 80.2 & \textbf{1366.4} 
    & 39.5 & \textbf{242.5} 
    & 100613.8 & \textbf{586171.2}\\
    
mjs-4 
    & 32.08\% & \textbf{36.97\%} 
    & 524 & \textbf{1870.9} 
    & 376.7 & \textbf{655.4} 
    & 569233.8 & \textbf{2273708} \\
    
\hline

\end{tabular}
\end{table*}
\renewcommand{\arraystretch}{1}

\subsection{Reported Bugs in Latest Versions of Applications}
\label{sec::eval::bugs}

\autoref{tab::reportedbugs} shows the summary of the reported bugs that were found by our tool, \topr, in the latest versions of \NumBuggyProjects projects. The table shows the number of bugs found per benchmark, the binary on which fuzzer was run along with the type, location, status, severity and the link of each reported bug.
Of the \NumBugs bugs found, \hdfBugs are in hdf5, \netcdfBugs in netcdf-c, \lrzipBugs in lrzip, \giflibBugs in giflib and \mjsBugs in mjs. \hdfBugsConfirmed bugs in hdf5, \netcdfBugs bugs in netcdf-c and \lrzipBugs bug in lrzip have all been confirmed by developers.
Furthermore, \hdfBugsHigh of the hdf5 bugs have been labelled as \textbf{``Priority - 1. High''} and \hdfBugsFixed of them have already been fixed.

\renewcommand{\arraystretch}{1.2}
\begin{table*}[!htbp]
  \centering
  \caption{Summary of the \NumBugs \emph{newly-discovered} bugs found by \topr. Details collected using ASAN.}
  
  \label{tab::reportedbugs}
  
\begin{tabular}{l cc ccccc}
    \hline
    \textbf{Benchmark} & \textbf{\#Bugs} & \textbf{Binary} & \textbf{Type} & \textbf{Location} & \textbf{Status} & \textbf{Severity} & \textbf{Bug Link} \\
    \hline
    
\rowcolor{gray!10} \multirow{12}{*}{} hdf5 & \hdfBugs & h5dump 
    & \hbof & H5Oginfo.c:104 & Fixed & High Priority & \href{https://github.com/HDFGroup/hdf5/issues/2601}{\underline{b1-link}} \\
    & & & \mallocerr & H5C.c:7120 & Fixed & High Priority & \href{https://github.com/HDFGroup/hdf5/issues/2600}{\underline{b2-link}} \\
    & & & \mleak & h5trav.c:124 & Fixed & High Priority & \href{https://github.com/HDFGroup/hdf5/issues/2598}{\underline{b3-link}} \\
    \rowcolor{gray!10} & & & \hbof & H5Olinfo.c:121 & Fixed & High Priority & \href{https://github.com/HDFGroup/hdf5/issues/2603}{\underline{b4-link}} \\
    & & & \hbof & H5Fint.c:2859 & Fixed &  High Priority & \href{https://github.com/HDFGroup/hdf5/issues/2604}{\underline{b5-link}} \\
    \rowcolor{gray!10} & & & \mleak & H5Oefl.c:137 & Fixed & High Priority & \href{https://github.com/HDFGroup/hdf5/issues/2605}{\underline{b6-link}} \\
    & & & \hbof & H5MM.c:311 & Fixed & High Priority & \href{https://github.com/HDFGroup/hdf5/issues/2606}{\underline{b7-link}} \\
    \rowcolor{gray!10} & & & \mleak & H5FL.c:237 & Fixed & High Priority & \href{https://github.com/HDFGroup/hdf5/issues/2599}{\underline{b8-link}} \\
    \rowcolor{gray!10} & & & \mleak & H5Opline.c:392 & Fixed & High Priority & \href{https://github.com/HDFGroup/hdf5/issues/2602}{\underline{b9-link}} \\
    & & & Out Of Memory & H5C.c:7120 & Fixed & High Priority & \href{https://github.com/HDFGroup/hdf5/issues/2607}{\underline{b10-link}} \\
    \rowcolor{gray!10} & & & \hbof & H5Oattr.c:144 & Fixed & High Priority & \href{https://github.com/HDFGroup/hdf5/issues/2529}{\underline{b11-link}} \\
    & & & \mleak & H5FL.c:246 & Not fixed & High Priority & \href{https://github.com/HDFGroup/hdf5/issues/2657}{\underline{b12-link}} \\
    \rowcolor{gray!10} & & & \segv & H5MM.c:311 & Fixed & - & \href{https://github.com/HDFGroup/hdf5/issues/2528}{\underline{b13-link}} \\

\hline

\multirow{5}{*}{} netcdf-c & \netcdfBugs & ncdump 
    & \segv & hdf5open.c:1333 & Confirmed & Pending & \href{https://github.com/Unidata/netcdf-c/issues/2664}{\underline{b14-link}} \\
    \rowcolor{gray!10} & & & \mleak & nchashmap.c:183 & Confirmed & Pending & \href{https://github.com/Unidata/netcdf-c/issues/2665}{\underline{b15-link}} \\
    & & & \hbof & hdf5open.c:1333 & Confirmed & Pending & \href{https://github.com/Unidata/netcdf-c/issues/2666}{\underline{b16-link}} \\
    \rowcolor{gray!10} & & & \hbof & hdf5var.c:2102 & Confirmed & Pending & \href{https://github.com/Unidata/netcdf-c/issues/2668}{\underline{b17-link}} \\
    & & & \mcpo & hdf5open.c:1333 & Confirmed & Pending & \href{https://github.com/Unidata/netcdf-c/issues/2667}{\underline{b18-link}} \\

\hline

\rowcolor{gray!10} lrzip-1 & \lrzipBugs & lrzip 
    & \mallocerr & stream.c:1692 & Confirmed & Pending & \href{https://github.com/ckolivas/lrzip/issues/242}{\underline{b19-link}} \\

\hline

\multirow{4}{*}{} giflib-1 & \giflibBugs & gifsponge 
    & \mleak & gifalloc.c:329 & Reported & Pending & \href{https://sourceforge.net/p/giflib/bugs/162/}{\underline{b20-link}} \\
    \rowcolor{gray!10} & & & \mleak & egif\_lib.c:101 & Reported & Pending & \href{https://sourceforge.net/p/giflib/bugs/161/}{\underline{b21-link}} \\
    & & & \mleak & gifalloc.c:331 & Reported & Pending & \href{https://sourceforge.net/p/giflib/bugs/163/}{\underline{b22-link}} \\
    \rowcolor{gray!10} & & & \mleak & gifalloc.c:53 & Reported & Pending & \href{https://sourceforge.net/p/giflib/bugs/164/}{\underline{b23-link}} \\

\hline

mjs-1 & \mjsBugs & mjs-bin & \segv & Unspecified & Reported & Pending & \href{https://github.com/cesanta/mjs/issues/242}{\underline{b24-link}} \\
\hline

\end{tabular}
\end{table*}
\renewcommand{\arraystretch}{1}

\section{Related Work}
\label{sec::rwork}

Recently, directed fuzzing has gained much popularity~\cite{beacon,selectfuzz,beak,gfuzz,rlf,switchfuzz,rdgfuzz,mc2,leofuzz,bugsbunny}.
Beacon~\cite{beacon} uses path-condition satisfiability to prune execution paths at runtime which incurs a high analysis overhead of approximately 20.8\%.
Prior work~\cite{sievefuzz} has already demonstrated that Beacon's runtime pruning suffers from the following drawbacks --- \ci its significant high cost reduces fuzzing throughput and \cii its imprecise control flow analysis tends to over-prune and aggressively remove reachable paths.
Contrary to this, \ci \topr employs an extremely light-weight static pruning mechanism that prunes code at compile time with an analysis overhead in the order of a few seconds, incurring no runtime overhead and \cii \topr's precise indirect flow analysis recovers indirect edges \wrt targets accurately.

SelectFuzz~\cite{selectfuzz} proposes a new metric for distance minimization that aids in the selection and exploration of paths specific to targets.
Beak~\cite{beak} is a hybrid directed fuzzer for smart contracts that also uses symbolic execution to improve target specific testing.
G-fuzz~\cite{gfuzz} is a directed fuzzer for Google's gVisor kernel that incorporates a combination of fine-grained distance calculation, target related syscall inference and utilization along with a 
dynamic switch between exploration and exploitation to enhance directed fuzzing.
RLF~\cite{rlf} utilizes deep reinforcement learning along with computing the distance between the path and the target to optimize fuzzing.
SwitchFuzz~\cite{switchfuzz} aims to address the issue of path exploration falling into a
local optimum using runtime feedback. It switches to other possible paths when the distance of test cases to target does not decrease over a period of time.
RDGFuzz~\cite{rdgfuzz} proposes to optimize \aflgo's seed energy allocation algorithm for better directed fuzzing.
LeoFuzz~\cite{leofuzz} proposes an energy scheduling strategy based on several relations between seeds and target locations and a strategy for better coordination of exploration, exploitation stages.
MC2~\cite{mc2} introduces a complexity-theoretic framework to convert directed greybox fuzzing into an oracle-guided search problem where the querying oracle receives feedback on how close an input is to the target.
BugsBunny~\cite{bugsbunny} aims to optimize directed fuzzing for complex hardware designs by proposing a distance-to-target feedback metric to minimize distance to target locations. 
Distinct from \topr, these techniques do \emph{not} use pruning to optimize directed fuzzing.

There has also been significant research effort towards making directed fuzzing more effective on complex\Comment{ user-space} programs including hybrid techniques that leverage symbolic execution.
Next, we discuss directed fuzzing and taint-based fuzzing techniques that include both greybox and whitebox fuzzers. 
Many coverage-based fuzzing techniques, both greybox and whitebox, have endeavored to increase code coverage while generating inputs~\cite{collafl,fairfuzz,tfuzz}.
Coverage-based greybox fuzzers~\cite{aflfast,vuzz,gbf3,afl,gbf5} primarily make use of light-weight instrumentation to increase code coverage.
AFL's~\cite{afl} instrumentation tracks basic block transitions and branch-taken hit counts and uses it to decide which of the generated inputs to preserve for fuzzing.
AFLFast~\cite{aflfast} investigates the possibility of exercising low-frequency paths to cover more paths in a given time unit.
Vuzzer~\cite{vuzz} triages basic blocks that select certain paths likely leading to a vulnerability target location.
The technique by Sparks et al.~\cite{gbf3} puts forward a genetic algorithm to increase code coverage and penetration depth into control-flow logic.

Symbolic execution based whitebox fuzzers are mostly implemented on symbolic execution engine, KLEE~\cite{kleepap}.
Most of the current directed fuzzers~\cite{df1, df2, df3, dse, df5, df6, df7, df8} exploit the systematic exploration of path, a principal characteristic of symbolic execution. Symbolic execution runs a program by assigning both concrete and symbolic values to variables\Comment{~\cite{se1,se2,se3,se4}}. 
Theoretically, symbolic execution would explore all possible paths in a program by forking at the conditional branches. 
However, in spite of being a widely used technique in testing, symbolic execution suffers from a few challenges that limit its usage in large real-world programs~\cite{seissues}. Chief among these is path explosion that prevents the technique from reaching deep parts of program code.

There has also been work in the area of directed symbolic execution. DiSE~\cite{dise} combines static analysis techniques for computing program differences between two versions to improve symbolic execution for regression testing. 
Fitnex~\cite{fitnex} uses target-specific distance computation to guide path exploration in symbolic execution.
eXpress~\cite{express} uses path pruning to enhance directed symbolic execution.

Apart from these, efforts combining both these strategies---fuzzing and symbolic execution---have also been attempted~\cite{fsym1,fsym2} in the past.
Also, there are many taint-based directed whitebox fuzzing techniques~\cite{buzzfuzz,vuzz,tdf} that dispense with the not so lightweight machinery of symbolic execution.
Buzzfuzz~\cite{buzzfuzz} is a dynamic whitebox technique that uses automatic taint tracking to mark the original input files at strategic critical locations.
Vuzzer~\cite{vuzz} (already mentioned above) is a coverage directed greybox fuzzer that also uses taint analysis to exercise not easily accessible code.

\section{Conclusion}
\label{sec::concl}

\topr is designed to enable fast and precise directed fuzzing.
Our evaluation demonstrates that \topr outperforms both \aflgo and \sfuzz in terms of \ci bug discovery (both number and speed), \cii target reachability and \ciii overall speed by achieving a higher test case throughput.
\topr found \NumBugs bugs in the latest versions of \NumBuggyProjects projects, of which \NumConfirmedBugs bugs have been already confirmed and \NumFixedBugs bugs fixed by the developers with \hdfBugsHigh bugs labelled as "Priority - 1. High"

\section*{Acknowledgment}
This material is based upon work supported by the National Science Foundation (NSF) under Grants NSF CCF-1763788, NSF CCF-1956374 and the Office of Naval Research (ONR) under Contract N68335-17-C-0558.

\bibliographystyle{plainurl}
\bibliography{11-refs,12-refs}

\end{document}